\documentclass[runningheads]{llncs}
\usepackage{version}

\includeversion{rep}
\excludeversion{conf}

\usepackage[hyphens]{url}
\usepackage[T1]{fontenc}
\usepackage{graphicx}
\usepackage{hyperref}
\usepackage{makecell}
\usepackage{listings}
\usepackage{xcolor}
\usepackage{bm}
\usepackage{orcidlink}
\usepackage{lineno}
\usepackage{prftree}
\newcommand{\dmd}{\ensuremath{\,\diamond\,}}
\newcommand\kickoff{departure}
\newcommand\landing{arrival}

\newcommand{\pt}{\phantom{0}}

\lstdefinelanguage{Lean}{
  morekeywords={
    import, class, where, abbrev, theorem, have, let, show, by,
    intros, calc, exact, nth_rw, infix, duper
  },
  sensitive=true,
  morecomment=[l]{--},
  morestring=[b]"
}

\hypersetup{
   a4paper,
   pdftex,
   pdfkeywords={},
   pdfborder={0 0 0},
   draft=false,
   bookmarksnumbered,
   bookmarks,
   bookmarksdepth=2,
   bookmarksopenlevel=2,
   bookmarksopen
}
\usepackage{amsfonts}
\usepackage{tikz}
\usepackage{array}
\usetikzlibrary{shapes, arrows.meta, positioning}
\usepackage{amsmath}
\usepackage{algorithm}
\usepackage{algorithmicx}
\usepackage[noend]{algpseudocode}
\usepackage{color}
\usepackage{cite}
\usepackage{multirow}
\usepackage{booktabs}
\usepackage{siunitx}
\usepackage{wasysym}

\newrobustcmd\B{\DeclareFontSeriesDefault[rm]{bf}{b}\bfseries}
\newcommand\sym[1]{\mathsf{#1}}

\spnewtheorem{examplex}[algorithm]{Example}{\bfseries}{\rmfamily}
\spnewtheorem{definitionx}[algorithm]{Definition}{\bfseries}{\rmfamily}

\begin{document}

\title{Tao's Equational Proof Challenge Accepted\begin{rep} (Technical~Report)\end{rep}}

\author{Lydia Kondylidou\inst{1}\orcidlink{0009-0001-9875-2627} \and
Jasmin Blanchette\inst{1}\orcidlink{0000-0002-8367-0936} \and
Marijn J.H. Heule\inst{2}\orcidlink{0000-0002-5587-8801}}
\authorrunning{L. Kondylidou et al.}
\institute{Ludwig-Maximilians-Universität München, Munich, Germany\\
\email{\{l.kondylidou,jasmin.blanchette\}@lmu.de} \and Carnegie Mellon University,
Pittsburgh, United States\\
\email{marijn@cmu.edu}}
\maketitle
\begin{abstract}
In the context of the Equational Theories Project,
Terence Tao posed the challenge of finding alternatives to a
complicated 62-step proof found by the Vampire superposition prover. We
introduce a proof minimization tool called Krympa.
Using a combination of brute force and heuristics, and exploiting both Vampire
and the Twee equational prover, the tool reduces the 62-step proof to 20~steps, each
corresponding to a rewrite.
In an empirical evaluation, it also performs well on 1431 equational
problems originating from the same project, reducing in particular a 151-step
proof to only 10~steps.

\keywords{Theorem provers \and Equational logic \and Proof minimization.}
\end{abstract}

\section{Introduction}
\label{sec:introduction}

The Equational Theories Project \cite{Bolan_The_Equational_Theories_2025},
launched in September 2024 by Fields med\-al\-ist Terence Tao, aims at exploring
the relations between different equational theories of magmas. A \emph{magma} is
a basic algebraic structure consisting of a set equipped with a single binary
operation $\diamond$ closed on that set. The project's first phase, concluded in
April 2025, focused on equational laws for magmas that contain at
most four applications of $\diamond$.

The project uses the Lean \cite{lean} proof assistant to express proofs and
counter\-examples but depends on automatic theorem provers and other
external tools. The problems explored in the project's first phase all fall
within first-order logic's unit equality fragment: They consist of a
universally quantified equation as the sole axiom and a universally quantified
equation as the proof goal, or conjecture.

For the problem $650\Longrightarrow448$, where 650 denotes the axiom $\forall
x,y,z.\; x = x \dmd (y \dmd ((z \dmd x) \dmd y))$ and 448 denotes the conjecture $\forall
x,y,z.\; x = x \dmd (y \dmd (z \dmd (x \dmd z)))$, the Vampire~\cite{vampire}
super\-posi\-tion prover found a particularly complex proof, with 62 inference
steps, excluding clausification and Skolemization.
Given that the proof is unintelligible,
Tao challenged the community to find ``an alternate proof, by
whatever means you wish---human, semi-automated, or automated''
\cite{lean-zulip}.

One idea was to run a specialized equational prover, Twee \cite{twee},
instead of Vampire, but this results in a very long, 137-step proof.
Another approach would be to use Lean's automation, such as the \texttt{aesop} \cite{aesop}, \texttt{canonical}
\cite{canonical}, \texttt{duper} \cite{duper}, and \texttt{grind}
\cite{lean-grind-tactic} tactics and LeanHammer \cite{zhu2025leanhammer}, to reconstruct
and compress consecutive superposition steps, in the style of Sledgehammer's
structured proof\pagebreak[2] \hbox{reconstruction}
\cite[Sect.~6.3]{semi-intelligible-blanchette-et-al}. This would yield a shorter
and more high-level proof, in which each step may combine multiple rewrites.
Our approach is orthogonal. Our working hypothesis is that the 62-step Vampire
proof, which emerged as the byproduct of a saturation process, is likely
suboptimal. By mixing and matching proofs generated by different
automatic provers, as proposed by Sutcliffe et
al.~\cite{Sutcliffe2011Combining}, we hoped to achieve a shorter, simpler
proof.

We introduce Krympa, a tool that minimizes equational proofs by
decomposing them into independently provable parts and reassembling them
into more concise, intelligible proofs. Specifically, starting from a
Vampire-generated proof, the tool transforms it into a direct proof
(Sect.~\ref{sec:proof-redirection}) and analyzes its inferences to break it
down into intermediate results that serve as candidate lemmas. Each of these lemmas
is then proved independently using Vampire and Twee
(Sect.~\ref{sec:subproof-generation}), the two leading systems in the unit
equality division of the 2025 edition of CASC \cite{Sut25-CASC}. The resulting proofs are then
combined into a single proof using heuristics
that favor shorter derivations
(Sect.~\ref{sec:subproof-combination}).

Given the 62-step Vampire proof of $650\Longrightarrow448$, our tool produces a
20-step proof, where 13~steps are generated by Twee and 7~steps
are generated by Vampire (Sect.~\ref{sec:application-to-taos-challenge}). In a larger
empirical evaluation, we applied the tool to 1431 provable implications from
the Equational Theories Project and obtained positive results
(Sect.~\ref{sec:experiments-on-other-equational-proofs}). In particular, the
tool reduced a 151-step Vampire proof to 10 steps.

Our tool is implemented in Rust, OCaml, and Python. Its source code is available at
\url{https://github.com/kondylidou/Krympa}. The files associated with Tao's
challenge and our empirical evaluation data are also available online
\cite{raw_data}.

\begin{conf}
An extended version of this paper is available online as a technical report at
\url{https://arxiv.org/TODO}.
\end{conf}

\section{Background}
\label{sec:background}

We briefly review the Vampire and Twee \begin{rep}automatic \end{rep}provers and their
\begin{rep}respective \end{rep}proof formats.

\subsection{Vampire and Superposition Proofs}

Vampire is a saturation-based theorem prover for first-order logic with equality
based on the superposition calculus \cite{10.1007/BFb0054258}. It implements highly
optimized search strategies and data structures, and integrates techniques such
as literal selection, term orders, redundancy elimination, strategy
scheduling, and portfolios.

Superposition works on implicitly universally quantified clauses. A
preprocessor performs clausification and Skolemization. For example, the axiom
$\forall x.\; \sym{f}(x) = \sym{g}(x)$ is transformed into $\sym{f}(x) =
\sym{g}(x)$, where $x$ is a free variable, and the conjecture $\forall x.\;
\sym{f}(x) = \sym{g}(x)$ is negated and transformed into $\sym{f}(\sym{sk})
\not= \sym{g}(\sym{sk})$, where $\sym{sk}$ is a Skolem constant. The objective is
to derive the contradictory clause $\bot$. For the unit
equality fragment, the calculus's two relevant inference rules are as follows:
\[
\prftree[r]{equality resolution}
{\strut t \not= u}
{\strut \bot}
\qquad
\prftree[r]{superposition}
{\strut t = t'}
{\strut s[u] \mathbin{\Bowtie} s'}
{\strut \mu(s[t'] \mathbin{\Bowtie} s')}
\]
The equality resolution rule has one premise, $t \not= u$, one conclusion,
$\bot$, and one side condition:\ that $t$ and $u$ are
unifiable.
The superposition rule has two premises
and one conclusion.
The $\bowtie$ symbol denotes either $=$ or $\not=$ throughout the rule.
The $=$ and $\not=$ operators are
commutative; for example, the premise $t \not= u$ can match
the disequation $\sym{f}(\sym{a}) \not= x$ either as is or as
$x \not= \sym{f}(\sym{a})$.
The premises are assumed to have mutually disjoint sets of variables, which can be
achieved by renaming.
The notation $s[\,]$ represents a term with a hole,
the terms $s[u]$ and $s[t']$ are obtained by filling the hole in $s[\,]$
with $u$ and $t'$,
and $\mu$ is a most general unifier of $t$ and $u$.
For example, the most general unifier of the terms $\sym{h}(\sym{a}, y)$
and $\sym{h}(x, \sym{b})$ is $\{x \mapsto \sym{a}{,}~ y \mapsto \sym{b}\}$;
applying it on both terms yields $\sym{h}(\sym{a}, \sym{b})$.
Finally, the rule has further side conditions, not shown here, that restrict
the search space.

\begin{examplex}
A subtle case of the superposition rule arises when both premises are the same
clause. In the following, the variable in
the second premise has been renamed to avoid a clash:
\[
\prftree[r]{superposition}
{\strut \sym{f}(\sym{f}(x)) = \sym{g}(x)}
{\strut \sym{f}(\sym{f}(x')) = \sym{g}(x')}
{\strut \sym{f}(\sym{g}(x)) = \sym{g}(\sym{f}(x))}
\]
This instance is obtained by taking
$t := \sym{f}(\sym{f}(x))$,
$t' := \sym{g}(x)$,
${\bowtie} := {=}$,
$s[\,] := \sym{f}([\,])$,
$u := \sym{f}(x')$,
$s' := \sym{g}(x')$,
and $\mu = \{x' \mapsto \sym{f}(x)\}$.
Applying the unifier $\mu$ to both premises yields the equations
$\sym{f}(\sym{f}(x)) = \sym{g}(x)$ and
$\sym{f}(\sym{f}(\sym{f}(x))) = \sym{g}(\sym{f}(x))$.
The inference replaces the subterm $\sym{f}(\sym{f}(x))$ in the second equation
with $\sym{g}(x)$ using the first equation as a left-to-right rewrite rule, and
derives the conclusion.\hfill\rule{0.7em}{0.7em}
\end{examplex}

\begin{examplex}
Vampire implements \emph{parallel superposition}, a variant of the superposition
rule in which all subterms that match a term are replaced.
\begin{conf}For example:\end{conf}
\begin{rep}The following inference illustrates this:\end{rep}
\[
\prftree[r]{parallel superposition}
{\strut \sym{b} = \sym{a}}
{\strut \sym{h}(\sym{b},\sym{a},\sym{b}) \not= \sym{h}(\sym{a},\sym{b},\sym{a})}
{\strut \sym{h}(\sym{a},\sym{a},\sym{a}) \not= \sym{h}(\sym{a},\sym{a},\sym{a})}
\]

\vspace{-\belowdisplayskip}
\vspace{-\baselineskip}
\noindent\hbox{}\hfill\rule{0.7em}{0.7em}
\end{examplex}

Superposition proofs are expressed in a linear format. They are refutational
and show how to derive~$\bot$ from the input axioms and the negated conjecture.

\begin{examplex}
\label{ex:sup-proof}
The following is a superposition proof from clauses 1--3:
\begin{quote}
\noindent
\begin{tabular}{@{}l@{}l@{\qquad}l@{}}
1.~ & $\sym{a} = \sym{b}$
  & {axiom} \\
2.~ & $\sym{f}(x) = x$
  & {axiom} \\
3.~ & $\sym{h}(\sym{f}(\sym{b}),\sym{a}) \not= \sym{h}(\sym{a},\sym{f}(\sym{b}))$
  & {negated conjecture} \\
4.~ & $\sym{h}(\sym{b},\sym{a}) \not= \sym{h}(\sym{a},\sym{b})$
  & {by parallel superposition from 2 and 3} \\
5.~ & $\sym{h}(\sym{a},\sym{a}) \not= \sym{h}(\sym{a},\sym{a})$
  & {by parallel superposition from 1 and 4} \\
6.~ & $\bot$
  & {by equality resolution from 5}
\end{tabular} \\[-\baselineskip]
\noindent\hbox{}\hfill\rule{0.7em}{0.7em}\kern-\leftmargin
\end{quote}
\end{examplex}

\subsection{Twee and Structured Equational Chain Proofs}

Twee is an automatic prover specialized for equational reasoning. It is
based on the unfailing completion procedure \cite{BACHMAIR19891}, an extension
of Knuth--Bendix completion \cite{KnuthBendix1970}.
As with superposition,
quantifiers are eliminated by a preprocessor.
In the DISCOUNT \cite{DS94b} and Waldmeister tradition
\cite{BuchHillenbrand1996}, Twee's proofs are structured as
a sequence of
lemmas, where the lemmas and the conjecture are proved by chains of equalities.
Twee introduces lemmas if they are needed more than once.
\begin{rep}Twee proofs are arguably more readable than Vampire proofs.\end{rep}

\begin{examplex}
\label{ex:twee}
The following is a Twee-style proof of goal 1 from axioms 1 and 2:

\begin{quote}
\noindent
\begin{minipage}[t]{.5\textwidth}
Axiom 1: $\sym{a} = \sym{b}$

\medskip

Axiom 2: $\sym{f}(x) = x$

\medskip

Lemma 3: $\sym{f}(\sym{b}) = \sym{a}$ \\
Proof: \\
\noindent\phantom{$=$~}$\sym{f}(\sym{b})$ \\
$=$~$\{$ by axiom 1 right-to-left $\}$ \\
\noindent\phantom{$=$~}$\sym{f}(\sym{a})$ \\
$=$~$\{$ by axiom 2 $\}$ \\
\noindent\phantom{$=$~}$\sym{a}$
\end{minipage}
\begin{minipage}[t]{.5\textwidth}
Goal 1: $\sym{h}(\sym{f}(\sym{b}), \sym{a}) = \sym{h}(\sym{a}, \sym{f}(\sym{b}))$ \\
Proof: \\
\noindent\phantom{$=$~}$\sym{h}(\sym{f}(\sym{b}), \sym{a})$ \\
$=$~$\{$ by lemma 3 $\}$ \\
\noindent\phantom{$=$~}$\sym{h}(\sym{a}, \sym{a})$ \\
$=$~$\{$ by lemma 3 right-to-left $\}$ \\
\noindent\phantom{$=$~}$\sym{h}(\sym{a}, \sym{f}(\sym{b}))$

\medskip

\noindent\phantom{Axiom 1}

\medskip

\noindent\phantom{Axiom 1}\hfill\rule{0.7em}{0.7em}\kern\leftmargin
\end{minipage}
\end{quote}
\end{examplex}

\section{Conversion to Direct Proofs}
\label{sec:proof-redirection}

Vampire generates proofs by refutation, whereas our mix-and-match approach
requires direct proofs. To bridge this gap, we transform Vampire proofs into
direct proofs. In the following sections, we will always use direct proofs.

In equational logic, to produce a direct proof, we first introduce
existential quantifiers for Skolem constants and universal quantifiers for
variables. For example, $\sym{h}(x,\sym{sk}) \not= x$ is transformed
into $\exists z.\>\forall x.\; \sym{h}(x,z) \not= x$. Then we apply the
contrapositive to all inferences in which a premise and the conclusion are
disequations to obtain positive equations. Thus, the
inference
\[
\prftree[r]{superposition}
{\strut \sym{h}(\sym{a}, y) = \sym{b}}
{\strut \sym{h}(x,\sym{sk}) \not= x}
{\strut \sym{b} \not= \sym{a}}
\]
becomes
\[
\prftree[r]{}
{\strut \forall y.\; \sym{h}(\sym{a}, y) = \sym{b}}
{\strut \sym{b} = \sym{a}}
{\strut \forall z.\>\exists x.\; \sym{h}(x,z) = x}
\]
Notice that negating $\exists z.\> \forall x.\; \sym{h}(x,z) \not= x$
flips the quantifiers\begin{rep} and eliminates a double negation\end{rep}.

Equality resolution inferences from a premise $t \not= t$ are
omitted since their con\-tra\-positives derive tautologies $t = t$.
This reduces proof length by~one.

\begin{examplex}
\label{ex:vampire-direct}
From Example~\ref{ex:sup-proof}'s proof by refutation, we obtain
this direct proof:
\begin{quote}
\noindent
\begin{tabular}{@{}l@{}l@{\qquad}l@{}}
1.~ & $\sym{a} = \sym{b}$
  & {axiom} \\
2.~ & $\forall x.\; \sym{f}(x) = x$
  & {axiom} \\
3.~ & $\sym{h}(\sym{b},\sym{a}) = \sym{h}(\sym{a},\sym{b})$
  & {from 1 and the tautology $\sym{h}(\sym{a},\sym{a}) = \sym{h}(\sym{a},\sym{a})$} \\
4.~ & $\sym{h}(\sym{f}(\sym{b}),\sym{a}) = \sym{h}(\sym{a},\sym{f}(\sym{b}))$
  & {from 2 and 3}
\end{tabular} \\[-\baselineskip]
\noindent\hbox{}\hfill\rule{0.7em}{0.7em}\kern-\leftmargin
\end{quote}
\end{examplex}

We will refer to the statements of proof steps such as 3 and 4 above as
\emph{lemmas}.
We will say that a
lemma $\ell$ is a \emph{direct dependency} of a lemma $\ell'$ if $\ell$ is
used in the proof of $\ell'$,
and that $\ell$ is an \emph{indirect dependency}, or
simply \emph{dependency}, of $\ell'$
if it is transitively a dependency of $\ell'$.

\section{Proof Generation for Lemmas}
\label{sec:subproof-generation}

Our approach starts by translating the main theorem into a TPTP
\cite{Sut24} input problem
and running Vampire to produce a baseline proof. This proof is turned\pagebreak[2] into
a direct proof. For each lemma in the direct proof, we
generate corresponding problems, with the objective of proving them using
Vampire and Twee. Three problem variants are generated:
\begin{enumerate}
\item \emph{Big-step problems} contain the axioms together with the lemma as the
conjecture, and nothing else. This allows us to investigate whether a radically
new proof, with different intermediate steps, can be found.

\smallskip
\item \emph{Small-step problems} contain the axioms together with the lemma as
the conjecture, and all lemmas derived prior to this lemma in the baseline proof as
additional axioms. This allows us to investigate whether a somewhat similar
variant of the original derivation can be found.

\smallskip
\item\emph{Abstracted problems} are variants of big-step problems that contain
the axioms together with an abstracted, possibly unprovable version of the
lemma as the conjecture. Specifically, selected subterms of the lemma---for
example, terms such as $x \dmd y$ that do not contain nested
applications---are replaced by fresh
variables. This
allows us to investigate whether a more general version of the lemma is
provable, ideally with a shorter, more abstract proof.
\end{enumerate}
Each problem is submitted to the two provers.
If a proof is found for a small-step problem, we expand it to recursively
include the shortest proofs of the lemmas used as axioms for the axioms
referenced in the proof. Ties are broken arbitrarily.

Next, we compare the proofs of the three problem variants corresponding to
the same lemma. If the abstracted problem has the shortest proof, the
lemma it proves is replaced in all small-step problems where it appears as
an axiom with the generalized lemma from the abstracted problem. Each updated
small-step problem is then re-proved, and if the result has fewer steps,
we replace the small-step problem's proof with it.

The length of a
Vampire-generated proof is the number of steps of its direct proof,
excluding preprocessing. For Twee, the length of a proof is the cumulative
number of equalities in the equality chains. Thus, the Vampire proof in
Example~\ref{ex:vampire-direct} has two steps, and the Twee proof in
Example~\ref{ex:twee} has four steps.

\begin{rep}
\begin{examplex}
Consider the first-order problem consisting of the axiom
\begingroup
\begin{align*}
    &\forall x,y,z.\;x = y \dmd (x \dmd (z \dmd (z \dmd z)))
\end{align*}
\endgroup
and the conjecture
\begingroup
\begin{align*}
    &\forall x,y,z.\;x = ((y \dmd (z \dmd y)) \dmd x) \dmd y.
\end{align*}
\endgroup
The first lemma, derived by a superposition inference
in which both premises are the axiom, is
\begingroup
\begin{align}
  &\forall x,y,z,w.\;w = z \dmd (w \dmd ((x \dmd (y \dmd (y \dmd y))) \dmd x)) \tag{1}.
\end{align}
\endgroup
The second lemma, derived by superposition from the axiom and
lemma~(1),~is
\begingroup
\begin{align}
  &\forall x,y,z.\; z = y \dmd (y \dmd x) \tag{2}.
\end{align}
\endgroup
We create big-step, small-step, and abstracted problems corresponding to these
two lemmas and give them to Vampire and Twee.
The big-step problems consist only of the axiom and the
corresponding lemmas as conjectures. The small-step problem for lemma~(1) is
identical to its big-step problem, since there are no previously derived
lemmas. By contrast, the small-step problem for lemma~(2) includes the axiom,
lemma~(1) as an additional axiom, and lemma~(2) as the conjecture. The abstracted
problem for lemma~(1) contains the axiom and the conjecture $\forall
x,y,z,w,v.\;w = z \dmd (w \dmd ((x \dmd (y \dmd v)) \dmd x))$, which was
obtained by replacing the subterm $y \dmd y$ in lemma~(1) by the
universally quantified variable $v$. Similarly, the abstracted problem for
lemma~(2) contains the axiom and the conjecture $\forall x,y,z,w.\; z = y \dmd
w$, which was obtained by replacing the subterm $y \dmd x$ in lemma~(2) by
the universally quantified variable~$w$.
\end{examplex}
\end{rep}

\section{Proof Construction for the Main Theorem}
\label{sec:subproof-combination}

Based on the lemmas' proofs generated in the previous phase,
our approach constructs a proof of the main theorem. The proof generally consists of
three segments. The first segment starts with the axioms and ends with the
derivation of a so-called \emph{\kickoff{} lemma}. The second segment derives a
so-called \emph{\landing{} lemma}. The third segment derives the conjecture.
Different candidates are considered as the \kickoff{} and \landing{}
lemmas, yielding different proofs. The proof with the fewest steps is chosen.

Specifically,
we first identify up to six lemmas that arise close to the end of
the baseline proof, including the conjecture, and consider them as candidate\pagebreak[2] \landing{} lemmas.
For each candidate \landing{} lemma, we
consider its dependencies as candidate \kickoff{} lemmas.
Then, for each candidate \kickoff{} lemma, we construct a problem with the
axioms and the \kickoff{} lemma's dependencies as the axioms and the \kickoff{}
lemma itself as the conjecture. We run both provers and, if at least one
succeeds, we use the shorter result as the proof of the first segment, unless an
even shorter proof was generated in the previous phase.

Next, for each pair of candidate \kickoff{} and \landing{} lemmas, we generate a new
problem with the original axioms, the \kickoff{} lemma, and its dependencies as
axioms and the \landing{} lemma as the conjecture. We run both provers and, if at
least one succeeds, we use the shorter result as the proof of the second
segment, unless an even shorter proof was generated earlier.
Finally, we generate a new problem with the original axioms, the
\kickoff{} lemma, its dependencies, and the \landing{} lemma as axioms and the
original conjecture as the conjecture. We run both provers and, if at least one
succeeds, we use the shorter result as the proof of the third segment, unless
an even shorter proof was generated earlier.

Without the segmentation, proof minimization could be intractable due to
combinatorial explosion. We chose to work with three segments, and six
candidate arrival lemmas, as a trade-off between performance and flexibility.
The number of candidate arrival lemmas is a parameter that can easily be
adjusted\begin{rep} if desired\end{rep}.

\begin{examplex}
\label{ex:landing-kickoff-example}
Before we review the three-segment proof construction approach in detail, let us
look at an example. The following sketch represents a baseline seven-step
Vampire direct proof of a theorem $A \Longrightarrow C$,
where $A$ denotes the axiom and $L_1,\dots,L_6$ are the lemmas used to derive
the conjecture $C$:
\begin{quote}
\noindent
\rlap{$A$}\phantom{$L_0$}\qquad axiom \\
{$L_1$}\qquad from $A$ \\
{$L_2$}\qquad from $A$ and $L_1$ \\
{$L_3$}\qquad from $L_1$ and $L_2$ \\
{$L_4$}\qquad from $L_2$ and $L_3$ \\
{$L_5$}\qquad from $L_3$ and $L_4$ \\
{$L_6$}\qquad from $A$ and $L_5$ \\
\rlap{$C$}\phantom{$L_0$}\qquad from $L_5$ and $L_6$
\end{quote}

In the first phase, for each lemma $L_1, \dots, L_6$ and the conjecture $C$,
we construct big-step, small-step, and
abstracted problems and try to prove them using Vampire and Twee, retaining
the shortest proof for each lemma. Suppose the following: The shortest proof of
$L_1$ has one step and is obtained from its big-step problem using Vampire;
for $L_2$ and $L_3$,
the shortest proofs are obtained from their small-step problems using Twee; for
$L_4$, the shortest proof is obtained from its abstracted problem using Twee;
and for $L_5$, $L_6$, and $C$, the shortest proofs are obtained from their
small-step problems using Vampire.

In the next phase,
the last five lemmas, $L_2, \dots, L_6$, and the conjecture $C$ are considered as candidate
\landing{} lemmas. We focus on $L_6$. The proof below, found by Vampire
for $L_6$'s small-step problem, is the shortest proof for $L_6$:
\begin{quote}
\noindent
\rlap{$A$}\phantom{$L_0$}\qquad axiom \\
{$L_1$}\qquad from $A$ \\
{$L_2$}\qquad from $A$ and $L_1$
\pagebreak[2]
\\
{$L_3$}\qquad from $L_1$ and $L_2$ \\
{$L_4$}\qquad from $L_2$ and $L_3$ \\
{$L_5$}\qquad from $L_3$ and $L_4$ \\
{$L_6$}\qquad from $A$ and $L_5$
\end{quote}
This proof happens to be identical to the first six steps of the baseline proof,
but in general it could be different.

Next, lemmas $L_1$ to $L_5$ are considered as candidate \kickoff{}
lemmas. We focus on~$L_3$.
The proof of conjecture $C$ is constructed by concatenating three segments. For the
first segment, we create a new problem with $A$, $L_1$, and $L_2$ as axioms,
since they are dependencies of the \kickoff{} lemma $L_3$ in the above proof of $L_6$,
and $L_3$ as the conjecture. We run both provers on this problem and obtain a
two-step Vampire proof of $L_3$ from $A$, $L_1$, and a new lemma $L_2'$.
Since $L_1$ is treated as an axiom, we must include its proof to obtain a
complete proof of $L_3$. In the first phase, we found a one-step Vampire proof
of $L_1$ from the axiom $A$, so we use it. In summary, the proofs of $L_1$ and
$L_3$ form the first segment, which consists of one step for $L_1$ and two
steps for $L_3$.

For the second segment, we create a new problem with $A$, $L_1$, $L_2'$, and $L_3$
as axioms and the \landing{} lemma $L_6$ as the conjecture. We run both provers
on this problem and obtain a two-step Twee proof of $L_6$ from $L_1$ and $L_3$.
Together with the first segment, this yields a five-step proof of $L_6$. Since
this proof is shorter than the six-step proof of $L_6$ presented above, it is
used as the second segment.

For the third segment, we create a new problem with $A$, $L_1$, $L_2'$, the
\kickoff{} lemma $L_3$, and the \landing{} lemma $L_6$ as axioms and
$C$ as the conjecture. We run both provers on this problem and
obtain a two-step Twee proof of $C$ from $L_2'$ and $L_3$. Since
this proof does not use the \landing{} lemma $L_6$, the second segment is
excluded from the result. Concatenating the first and third segments yields a new
five-step proof of $C$, which is two steps shorter than the baseline proof:
\begin{quote}
\noindent
\rlap{$A$}\phantom{$L_0$}\qquad axiom \\
$L_1$\qquad from $A$ \\
$L_2'$\qquad from $A$ and $L_1$ \\
$L_3$\qquad from $L_1$ and $L_2'$ \\
\rlap{$C$}\phantom{$L_0$}\qquad by a two-step equality chain using $L_2'$ and $L_3$
\end{quote}

Finally, other combinations of candidate \kickoff{} and \landing{} lemmas are
considered, and the shortest proof is retained.
\hfill\rule{0.7em}{0.7em}\kern-\leftmargin
\end{examplex}

\subsection{Construction of the Dependency Graph}
\label{ssec:construction-of-the-dependency-graph}

We identify lemmas occurring close to the end of the derivation as
candidate {\landing{} lemmas}. Different candidates typically
depend on substantially different subsets of earlier lemmas.
Each candidate therefore induces its own dependency chain, and different choices can lead to
substantially different proof lengths.
We consider six candidate \landing{} lemmas extracted from the baseline
proof, including the conjecture itself, since our approach may produce a shorter
proof of the conjecture by reproving it directly from a minimized
dependency set.
All six candidates can be processed in parallel.

For every candidate, we build a dependency graph that captures the
lemmas required to derive it. Direct dependencies are determined from
the shortest Vampire or Twee proof obtained for each lemma. Here, shortest
means fewest inference steps, not fewest axioms.
Given that we generate three problem
variants and run two provers, up to six proofs per lemma are considered.
A lemma $\ell$ is considered to directly depend on a lemma
$\ell'$ if the
shortest proof of $\ell$ uses $\ell'$ as an axiom.
Thus, for big-step and
abstracted problems, only the original axioms can be dependencies.
For small-step problems, each step in a Vampire proof and each
lemma in a Twee proof is considered a lemma.

The dependency graph associated with a candidate \landing{} lemma is a
directed acyclic graph (DAG) whose
nodes correspond to lemmas and whose edges express derivability between them.
Formally, let $V$ be a finite set of lemmas, each represented by an equation
and a set of direct dependencies on other lemmas. We construct a DAG $(V,E)$,
where each vertex $\ell \in V$ corresponds to a lemma and each edge $(\ell,\ell')
\in E$ indicates a direct dependency of $\ell$ on~$\ell'$.
As an optimization, we merge lemmas that are identical up to the naming of
variables, keeping the shortest proof.

\kern-0.5pt

\subsection{Construction of the First Proof Segment}
\label{ssec:construction-of-the-first-segment}

For each candidate \landing{} lemma, we investigate whether all lemmas
included in its dependency graph are needed to derive it or whether a shorter
proof can be obtained by choosing a {\kickoff{} lemma} and recomputing parts of
the derivation by combining proofs generated by the provers.

As candidate \kickoff{} lemmas, we consider all lemmas in the DAG.
Let $\ell$ be a candidate \kickoff{} lemma. If $\ell$ depends only on the axioms,
we take the shortest big-step, small-step, or abstracted proof previously
found by Vampire or Twee. Otherwise,
we build a problem that includes $\ell$'s dependencies in the DAG as axioms and the
\kickoff{} lemma as the conjecture, and we run Vampire and Twee. If at least
one of them succeeds, we choose the shorter proof as $\ell$'s proof. This
derivation, together with the shortest proofs of $\ell$'s dependencies
generated for the big-step, small-step, or abstracted problems,
forms the first segment of the final proof. However, if we found an even
shorter proof for the big-step, small-step, or abstracted problem, we use that
proof instead. For small-step proofs, we must also include the proofs of the
lemmas encoded as axioms.

\subsection{Construction of the Remaining Proof Segments}

To construct the second segment, we generate a problem with the \kickoff{} lemma
and its dependencies as axioms and the \landing{} lemma as the conjecture, and
run both provers. If at least one of them succeeds, we choose the shorter
proof as the proof of the \landing{} lemma.
As above, we fall back on the proof of a big-step, small-step, or abstracted
problem if it is even shorter.

Finally, to construct the third segment, we generate a problem with the
\kickoff{} lemma, the \landing{} lemma, and their dependencies as axioms and the
original conjecture as the conjecture, and invoke both provers. If at least one of
them succeeds, we choose the shorter proof as the proof of the original conjecture. As
above, we fall back on a previously derived proof if it is even shorter.

The final proof is obtained by concatenating the three segments. When we
concatenate, lemmas from one segment are passed as axioms to subsequent
segments. Twee equality chains used to prove lemmas are not visible in
subsequent segments. The final proof might contain unreferenced lemmas; these are
pruned.

\begin{rep}
\begin{examplex}
As a concrete example, consider the theorem
\begingroup
\begin{align*}
    &(\forall x,y,z.\;x = ((x \dmd (y \dmd z)) \dmd z) \dmd z)
    \Longrightarrow
    \forall x,y,z.\;x = ((x \dmd y) \dmd (z \dmd x)) \dmd y.
\end{align*}
\endgroup
When applied to this problem, our tool first runs Vampire to obtain a
seven-step baseline proof. It then constructs seven problems of each variant (big-step,
small-step, and abstracted) and attempts to prove them using Vampire and Twee.
Among the candidate \landing{} lemmas, the shortest proof is found by selecting
\begingroup
\[\forall x,y,z.\;x \dmd y = x \dmd z. \tag{2}\]
\endgroup
Our tool then constructs the dependency graph for this lemma and, among the
candidate \kickoff{} lemmas, selects
\begingroup
\begin{align}
    &\forall x,y,z.\;x = ((x \dmd y) \dmd z) \dmd z. \tag{1}
\end{align}
\endgroup
\looseness=-1
Following the inference steps of the baseline Vampire proof, our
tool derives lemma~(1) by applying a superposition inference with the axiom
$x = ((x \dmd (y \dmd z)) \dmd z) \dmd z$
as the first premise and a renamed copy
$x' = ((x' \dmd (y' \dmd z')) \dmd z') \dmd z'$
as the second premise.
The most general unifier of the first premise's right-hand side and the subterm
$y' \dmd z'$ of the second premise is
$\{y' \mapsto (x \dmd (y \dmd z)) \dmd z{,}~
z' \mapsto z\}$. Applying the unifier to both premises yields the equations
$x = ((x \dmd (y \dmd z)) \dmd z) \dmd z$
and
$x' = ((x' \dmd (((x \dmd (y \dmd z)) \dmd z) \dmd z)) \dmd z) \dmd z$.
The superposition inference replaced the subterm $((x \dmd (y \dmd z)) \dmd z)
\dmd z$ in the second equation with $x$ using the first equation
as a right-to-left rewrite rule, and thus
derived lemma~1, up to the naming of variables.
Lemma~(2) is then proved using Twee from lemma~(1) and the axiom.
Finally, our tool proves the conjecture using Twee from lemmas (1)~and~(2).
The resulting proof has five steps, including Twee subproofs.
\end{examplex}
\end{rep}

\subsection{Proof Output}

Our tool generates the minimized proof in a native format,
from which two Lean outputs are produced.
\begin{rep}
Since our proofs are structured as direct chains of equalities with lemmas, Lean
is a more natural output format than TSTP \cite{Sut24}.
Lean proofs can also be checked by Lean's proof checker,
and they can be used to communicate with
the mathematicians working on the Equational Theories Project.

\end{rep}
The first Lean output is
a step-by-step formalization using the \verb|calc| tactic to reconstruct
equality chains. It applies the \texttt{duper} tactic to fill in the
subproofs. For example, a proof of $t_1 = t_2 = t_3 = t_4$ becomes
\begin{lstlisting}[basicstyle=\normalsize\ttfamily]
calc
  W = X := by duper …
  _ = Y := by duper …
  _ = Z := by duper …
\end{lstlisting}
where the ellipses stand for \verb|duper|'s arguments. The second Lean output is a more
compact formalization in which each lemma is proved directly using Lean's automation
without including the intermediate steps in equality chains.

\section{Application to Tao's Challenge}
\label{sec:application-to-taos-challenge}

We implemented our approach and tried the resulting tool, Krympa, on Tao's
challenge theorem $650\Longrightarrow448$:
\begingroup
\begin{align*}
  &(\forall x,y,z.\; x = x \dmd (y \dmd ((z \dmd x) \dmd y)))
  \Longrightarrow
  \forall x,y,z.\; x = x \dmd (y \dmd (z \dmd (x \dmd z))).
\end{align*}
\endgroup
Our tool first ran Vampire to obtain a baseline 62-step superposition proof.
Then it constructed 62 problems of each variant (big-step, small-step, and abstracted)
and tried to prove them using Vampire and Twee. Among the six candidate
\landing{} lemmas, the shortest proof was found by selecting
\begingroup
\begin{align}
  &\forall x,y,z.\;
  x = x \dmd
  {(\textcolor{blue}{(y \dmd ((z \dmd y) \dmd y))} \dmd x)}. \tag{lemma 9}
\end{align}
\endgroup
The coloring highlights repeating patterns. Next, our tool constructed
the dependency graph for this lemma. The DAG contained
37~lemmas. It was based on big- and small-step proofs.

Among the 10 candidate \kickoff{} lemmas, our tool found the shortest proof
by selecting
\begingroup
\begin{align}
  &\forall x,y,z,w.\;
  \textcolor{blue}{(x \dmd ((y \dmd x) \dmd x))} \dmd z = \notag \\[-\jot]
  &\quad (\textcolor{blue}{(x \dmd ((y \dmd x) \dmd x))} \dmd z) \dmd (w \dmd (\textcolor{blue}{(x \dmd ((y \dmd x) \dmd x))} \dmd w)). \tag{lemma 7}
\end{align}
\endgroup
According to the DAG, the shortest proof of this lemma was found by running
Vampire on the small-step problem consisting of the axiom and\pagebreak[2]
\begingroup
\begin{align}
  &\forall x,y,z,w.\;
  x \dmd (\textcolor{blue}{(y \dmd ((z \dmd y) \dmd y))} \dmd x) = \notag \\[-\jot]
  &\quad (x \dmd (\textcolor{blue}{(y \dmd ((z \dmd y) \dmd y))} \dmd x)) \dmd \notag \\[-\jot]
  &\quad \quad (w \dmd (\textcolor{blue}{(y \dmd ((z \dmd y) \dmd y))} \dmd w)). \tag{lemma 5}
\end{align}
\endgroup

The shortest proof of lemma~5 was found by running Vampire on the corresponding
big-step problem. The following lemmas were derived:
\begingroup
\begin{align}
  &\forall x,y,z,w.\;
  x \dmd ((y \dmd z) \dmd x) = \notag \\[-\jot]
  &\quad (x \dmd ((y \dmd z) \dmd x)) \dmd (w \dmd (z \dmd w)) \tag{lemma 1} \\
  &\forall x,y,z,w,v,u.\;
  x \dmd ((y \dmd ((z \dmd w) \dmd y)) \dmd x) = \notag \\[-\jot]
  &\quad (x \dmd ((y \dmd ((z \dmd w) \dmd y)) \dmd x)) \dmd (v \dmd ((u \dmd (w \dmd u)) \dmd v)) \tag{lemma 2} \\
  &\forall x,y,z,w,v.\;
  x \dmd (y \dmd x) = \notag \\[-\jot]
  &\quad (x \dmd (y \dmd x)) \dmd (z \dmd ((w \dmd ((v \dmd y) \dmd w)) \dmd z)) \tag{lemma 3} \\
  &\forall x,y,z,w,v.\;
  x \dmd (y \dmd x) = \notag \\[-\jot]
  &\quad (x \dmd (y \dmd x)) \dmd ((z \dmd (y \dmd z)) \dmd (w \dmd ((v \dmd y) \dmd w))) \tag{lemma 4}
\end{align}
\endgroup
Specifically, following the inference steps of the baseline Vampire proof, our tool derived
lemma~1 by applying a superposition inference with the axiom
$x = x \dmd (y \dmd ((z \dmd x) \dmd y))$
as the first premise and a renamed copy
$x' = x' \dmd (y' \dmd ((z' \dmd x') \dmd y'))$
as the second premise.
The most general unifier of the first premise's right-hand side and the subterm
$z' \dmd x'$ of the second premise is
$\{x' \mapsto y \dmd ((z \dmd x) \dmd y){,}~
z' \mapsto x\}$. Applying the unifier to both premises yields the equations
$x = x \dmd (y \dmd ((z \dmd x) \dmd y))$
and
$y \dmd ((z \dmd x) \dmd y) = (y \dmd ((z \dmd x) \dmd y)) \dmd (y' \dmd ((x \dmd
(y \dmd ((z \dmd x) \dmd y))) \dmd y'))$.
The superposition inference replaced the subterm $x \dmd (y \dmd ((z \dmd x)
\dmd y))$ in the second equation with $x$ using the first equation
as a right-to-left rewrite rule, and thus
derived lemma~1, up to the naming of variables. Lemmas 2~to~4 were derived
similarly following the steps of the baseline Vampire proof.

Next, from the axiom and lemma~5 our tool proved the \kickoff{} lemma (lemma~7)
using Vampire. In the proof,
\begingroup
\begin{align}
  &\forall x,y,z,w.\;
  \textcolor{blue}{(x \dmd ((y \dmd x) \dmd x))} \dmd z = \notag \\[-\jot]
  &\quad (\textcolor{blue}{(x \dmd ((y \dmd x) \dmd x))} \dmd z) \dmd ((w \dmd (\textcolor{blue}{(x \dmd ((y \dmd x) \dmd x))} \dmd w)) \dmd \notag \\[-\jot]
  &\quad \quad (z \dmd (\textcolor{blue}{(x \dmd ((y \dmd x) \dmd x))} \dmd z))) \tag{lemma 6}
\end{align}
\endgroup
was derived by applying a superposition inference with the axiom as the first
premise and lemma~5 as the second premise. Then, lemma~7 was derived by applying
a superposition inference with lemma~6 as the first premise and lemma~5 as the
second premise.

After, from the axiom and lemma~7, our tool proved the \landing{} lemma (lemma~9)
using Twee.
For this proof, Twee introduced the intermediate step
\begingroup
\begin{align}
  &\forall x,y,z,w.\;
  {\textcolor{blue}{(y \dmd ((z \dmd y) \dmd y))} \dmd w} = \notag \\[-\jot]
  &\quad (
  {\textcolor{blue}{(y \dmd ((z \dmd y) \dmd y))} \dmd w}) \dmd
  {(\textcolor{blue}{(y \dmd ((z \dmd y) \dmd y))} \dmd x)}. \tag{lemma 8}
\end{align}
\endgroup

Finally, assuming all the lemmas derived so far, our tool proved the conjecture
from lemmas~5 and~9 using Twee. The resulting proof has 20 steps,
including three Twee-generated chains of equalities.

Below we present the final proof adapted from our tool's detailed Lean output.
Instead of relying on proof automation, we use the \verb|nth_rw| tactic, which
performs a single rewrite step, where the numeric argument indicates which
matching occurrence should be rewritten. In one case, two numbers are supplied,
corresponding to a parallel rewrite.

\begin{lstlisting}
theorem Equation650_implies_Equation448 (G : Type _) [Magma G]
      (op_law : ∀ x y z : G, x = x◇(y◇((z◇x)◇y))) :
    ∀ x y z : G, x = x◇(y◇(z◇(x◇z))) :=
  have lemma1 (x y z w : G) :
      x◇((y◇z)◇x) = (x◇((y◇z)◇x))◇(w◇(z◇w)) := by
    nth_rw 3 [op_law z x y]
    exact op_law (x◇((y◇z)◇x)) w z

  have lemma2 (x y z w v u : G) :
      x◇((y◇((z◇w)◇y))◇x) =
      (x◇((y◇((z◇w)◇y))◇x))◇(v◇((u◇(w◇u))◇v)) := by
    nth_rw 1 2 [lemma1 y z w u]
    exact lemma1 x (y◇((z◇w)◇y)) (u◇(w◇u)) v

  have lemma3 (x y z w v : G) :
      x◇(y◇x) = (x◇(y◇x))◇(z◇((w◇((v◇y)◇w))◇z)) := by
    nth_rw 1 [lemma1 w v y x]
    exact op_law (x◇(y◇x)) z (w◇((v◇y)◇w))

  have lemma4 (x y z w v : G) :
      x◇(y◇x) = (x◇(y◇x))◇((z◇(y◇z))◇(w◇((v◇y)◇w))) := by
    nth_rw 1 [lemma1 w v y z]
    exact lemma3 x y (z◇(y◇z)) w v

  have lemma5 (x y z w : G) :
      x◇((y◇((z◇y)◇y))◇x) =
      (x◇((y◇((z◇y)◇y))◇x))◇(w◇((y◇((z◇y)◇y))◇w)) := by
    nth_rw 1 [lemma2 w y z y x ((z◇y)◇y)]
    exact lemma4 x (y◇((z◇y)◇y)) w x ((z◇y)◇y)

  have lemma6 (x y z w : G) :
      (x◇((y◇x)◇x))◇z =
      ((x◇((y◇x)◇x))◇z)◇((w◇((x◇((y◇x)◇x))◇w))◇
        (z◇((x◇((y◇x)◇x))◇z))) := by
    nth_rw 1 [lemma5 z x y w]
    exact op_law ((x◇((y◇x)◇x))◇z) (w◇((x◇((y◇x)◇x))◇w)) z

  have lemma7 (x y z w : G) :
      (x◇((y◇x)◇x))◇z =
      ((x◇((y◇x)◇x))◇z)◇(w◇((x◇((y◇x)◇x))◇w)) := by
    nth_rw 1 [lemma5 w x y z]
    exact lemma6 x y z w

  have lemma8 (x y z w : G) :
    ((x◇((y◇x)◇x))◇z)◇((x◇((y◇x)◇x))◇w) =
    (x◇((y◇x)◇x))◇z := by
    let T := x◇((y◇x)◇x)
    calc
      (T◇z)◇(T◇w) = 
      ((T◇z)◇((T◇w)◇((T◇(T◇w))◇((w◇(T◇w))◇(T◇(T◇w)))))) := by
        nth_rw 1 [←op_law]
      _ = ((T◇z)◇((T◇w)◇((T◇(T◇w))◇((w◇(T◇w))◇
            (T◇((T◇w)◇(w◇(T◇w)))))))) := by
        nth_rw 1 [←lemma7]
      _ = ((T◇z)◇((T◇w)◇((T◇(T◇w))◇((w◇(T◇w))◇
            ((T◇((T◇w)◇(w◇(T◇w))))◇(((T◇w)◇(w◇(T◇w)))◇
              (T◇((T◇w)◇(w◇(T◇w)))))))))) := by
        nth_rw 2 [←lemma7]
      _ = ((T◇z)◇((T◇w)◇((T◇(T◇w))◇(w◇(T◇w))))) := by
        nth_rw 1 [←op_law]
      _ = ((T◇z)◇((T◇w)◇(T◇(T◇w)))) := by
        nth_rw 1 [←lemma7]
      _ = ((x◇((y◇x)◇x))◇z) := by
        nth_rw 1 [←lemma7]

  have lemma9 (x y z : G) :
      (x◇((y◇((z◇y)◇y))◇x)) = x := by
    calc
      (x◇((y◇((z◇y)◇y))◇x)) =
      (x◇(((y◇((z◇y)◇y))◇x)◇((y◇((z◇y)◇y))◇x))) := by
        nth_rw 1 [lemma8]
      _ = (x◇(((y◇((z◇y)◇y))◇x)◇(((y◇((z◇y)◇y))◇x)◇
            ((y◇((z◇y)◇y))◇x)))) := by
        nth_rw 2 [lemma8]
      _ = x := by
        nth_rw 1 [←op_law]

  show _ by
    intros x y z
    calc
      x = x◇((x◇((y◇x)◇x))◇x) := by
        nth_rw 1 [lemma9]
      _ = (x◇((x◇((y◇x)◇x))◇x))◇((y◇(z◇(x◇z)))◇
            ((x◇((y◇x)◇x))◇(y◇(z◇(x◇z))))) := by
        nth_rw 1 [←lemma5]
      _ = x◇((y◇(z◇(x◇z)))◇((x◇((y◇x)◇x))◇
            (y◇(z◇(x◇z))))) := by
        nth_rw 1 [lemma9]
      _ = x◇(y◇(z◇(x◇z))) := by
        nth_rw 1 [lemma9]
\end{lstlisting}

\kern-2pt

\section{Experiments on Other Equational Proofs}
\label{sec:experiments-on-other-equational-proofs}

To assess the general potential of our approach, we evaluated our tool on a set
of equational theorems obtained from the Equational Theories Project repository
\cite{Bolan_The_Equational_Theories_2025}.
We selected all problems in the 13 Lean files
\verb|Proofs1| to \verb|Proofs13| that have a proof and translated them to
TPTP problem files, yielding 1431~benchmarks.

For each file, we invoked our tool's TPTP problem generator, which parses the
Lean theorems and produces corresponding TPTP problem files. For each problem,
our tool was given 600 seconds to produce a minimized proof using Vampire and Twee; on failure, the
baseline Vampire proof was output. A time limit of 10 seconds was used for each
prover invocation. The 600-second overall limit is generous; most
proofs are minimized in well under a minute.
The experiments were conducted on a server equipped with a dual-socket AMD EPYC
9965 system (384 cores, 768 threads) running at 2.25--3.70~GHz with 3~TiB of DDR5
ECC RAM, and running Debian GNU/Linux~13 (kernel 6.17.13+deb13-amd64).

Overall, proofs for the 13 Lean files have an average length of 6.6 steps
before minimization and 4.5 steps after minimization using the combination of
small-step and abstracted problems and both provers. This corresponds to a 31.5\%
decrease, showing that even short proofs can often be made shorter.

\begin{table}[b!]
\centering
\caption{Comparison of proof lengths before and after minimization for problems with baseline proofs of at least 15 steps}
\label{tab:proof_lengths_long}
\begin{tabular}{@{}l@{\kern1em}c@{\kern1em}c@{\kern1em}c@{\kern1em}c@{\kern1em}c@{\kern1em}c@{}}
\toprule
\relax{File} & \relax{Num.\ problems} & \relax{Avg.\ before} & \multicolumn{4}{c}{\relax{Avg.\ after}} \\
\cmidrule(lr){4-7}
 &  & & \relax{~~BA~~} & \relax{~~SA~~} & \relax{~~BS~~} & \relax{~~BSA~~} \\
\midrule
\verb|Proofs1|  & \pt9   & 17.3 & 16.0 & \textbf{13.1} & 13.3 & 13.3 \\
\verb|Proofs2|  & \pt8   & 16.9 & 14.4 & \textbf{11.5} & \textbf{11.5} & \textbf{11.5} \\
\verb|Proofs3|  & \pt7   & 19.3 & 15.6 & 10.9 & \textbf{10.7} & 10.9 \\
\verb|Proofs4|  & 11   & 19.1 & 14.3 & \textbf{10.9} & 11.4 & 11.4 \\
\verb|Proofs5|  & \pt9   & 20.1 & 17.6 & 12.8 & \textbf{11.9} & \textbf{11.9} \\
\verb|Proofs6|  & 11   & 25.6 & 18.9 & \textbf{12.5} & 12.6 & 12.6 \\
\verb|Proofs7|  & 11   & 37.2 & 19.7 & \textbf{11.8} & 11.9 & 11.9 \\
\verb|Proofs8|  & \pt7   & 24.4 & 15.6 & \textbf{12.3} & 13.1 & 13.1 \\
\verb|Proofs9|  & 14   & 39.8 & 29.0 & \textbf{13.1} & 14.1 & 14.1 \\
\verb|Proofs10| & \pt2   & 21.5 & 16.0 & \textbf{\pt8.0} & 11.0 & 11.0 \\
\verb|Proofs11| & \pt5   & 25.4 & 22.4 & \textbf{13.0} & 14.0 & 14.0 \\
\verb|Proofs12| & 13   & 24.6 & 16.5 & \textbf{\pt8.0} & \pt8.5 & \pt8.5 \\
\verb|Proofs13| & 10 & 35.3 & 27.7 & \textbf{\pt9.1} & 10.1 & 10.1 \\
\midrule
Total                  & \llap{1}17 & 26.3 & {19.2} & {\textbf{11.4}} & {11.8} & {11.8} \\
\bottomrule
\end{tabular}
\end{table}

Since longer proofs present more opportunities for minimization, we
now focus on the 117~benchmarks whose baseline proofs have at least 15~steps.
Table~\ref{tab:proof_lengths_long} compares proof lengths before and after
minimization.
The ``{Avg.\ before}'' column shows the average number of steps in
the baseline proofs. The ``{Avg.\ after}'' columns report the average proof
length after minimization under four configurations, which differ in which
problem variants are used:
``BA'' denotes the combination of the big-step and abstracted variants;
``SA'' denotes the combination of the small-step and abstracted variants;
``BS'' denotes the combination of the big- and small-step variants;
and ``BSA'' denotes the combination of all three variants.

The results show an often substantial reduction in proof length.
\relax{SA} generally yielded the shortest proofs.
Across all problems for the 13 Lean files, the average reduction with
\relax{SA} is 56.7\%.
\relax{BS} and \relax{BSA} also produced substantial reductions, whereas
\relax{BA} generally yielded the least improvements.

It might seem counterintuitive that SA, which does not consider big-step
problems,
outperforms BSA. However, the nonmonotonicity is to be expected. Provers are
nondeterministic, especially when invoked with a time limit; for example, a
prover might sometimes find a proof after 9.98~seconds and sometimes time out
after 10~seconds.
Moreover, our approach makes different heuristic choices when constructing the
three proof segments depending on which problem variants are used. As a result,
SA might find a short proof that escapes BSA.

The reduction in proof length is especially noticeable in individual cases.
The problem $2666\Longrightarrow3460$ has a Vampire proof
with 51 inference steps, which our tool reduces to only 12 single rewrite steps,
and $2923\Longrightarrow2628$ is
reduced from 180 steps to only 34.
The problem $3569\Longrightarrow3957$ is reduced from 92
to 23 steps and, even more dramatically, $3957\Longrightarrow3971$ is reduced
from 141 steps to only 23. Furthermore, $2860\Longrightarrow2660$ is
reduced from 44 to 14 steps, and $723\Longrightarrow872$ goes
from 57 to 13~steps. Finally, $947\Longrightarrow3897$
underwent the largest reduction, from 151 to 10 steps.
\begin{rep}Its Lean proof is shown below. \end{rep}
Overall, these results demonstrate that our
approach produces shorter proofs across a diverse set of equational theorems.

\begin{rep}
\begin{lstlisting}
theorem Equation947_implies_Equation3897 (G : Type _) [Magma G]
      (op_law : ∀ x y z : G, x = y◇((z◇x)◇(y◇x))) :
    ∀ x y z : G, x◇x = (y◇(z◇x))◇x :=
  have lemma1 (x y z w : G) :
      (x◇y)◇(z◇y) = w◇(y◇(w◇((x◇y)◇(z◇y)))) := by
    nth_rw 3 [op_law y z x]
    exact op_law ((x ◇ y) ◇ (z ◇ y)) w z

  have lemma2 (x y z : G) :
      (x◇y)◇(z◇y) = z◇(y◇y) := by
    nth_rw 4 [op_law y z x]
    exact lemma1 x y z z

  have lemma3 (x y : G) :
      y = x◇(x◇(y◇y)) := by
    nth_rw 1 [←lemma2 x y x]
    exact op_law y x x

  have lemma4 (x y z : G) :
      x◇x = (y◇(z◇x))◇x := by
    calc
      x◇x = (y◇(z◇x))◇((y◇(z◇x))◇((x◇x)◇(x◇x))) := by
        nth_rw 1 [←lemma3]
      _ = (y◇(z◇x))◇((y◇(z◇x))◇
            (z◇(z◇(((x◇x)◇(x◇x))◇((x◇x)◇(x◇x)))))) := by
        nth_rw 2 [←lemma3]
      _ = (y◇(z◇x))◇((y◇(z◇x))◇
            (z◇(z◇((x◇x)◇((x◇x)◇(x◇x)))))) := by
        nth_rw 1 [lemma2]
      _ = (y◇(z◇x))◇((y◇(z◇x))◇(z◇(z◇x))) := by
        nth_rw 1 [←lemma3]
      _ = (y◇(z◇x))◇(z◇((z◇x)◇(z◇x))) := by
        nth_rw 1 [lemma2]
      _ = (y◇(z◇x))◇(z◇(z◇(x◇x))) := by
        nth_rw 1 [lemma2]
      _ = (y◇(z◇x))◇x := by
        nth_rw 1 [←lemma3]

  show _ by
    exact lemma4
\end{lstlisting}
\end{rep}

\section{Related Work}
\label{sec:related-work}

At least two other researchers took on Tao's challenge. Kinyon \cite{kinyon}
found a 24-step proof (excluding preprocessing) of $650\Longrightarrow \forall
x, y.\; x = x \dmd y$ using Prover9 \cite{prover9},
from which $650\Longrightarrow448$ follows by instantiation.
Later, Le Floch \cite{le-floch} developed a pen-and-paper proof and translated
it to Lean. The Lean proof relies on only 14~rewrite steps but
includes additional reasoning as proof terms, and two of the rewrite steps are
parallel, so the overall length is similar to ours. The proof is ``loosely
based'' on the output of multiple Prover9 runs ``with
intermediate results thrown in as assumptions or as goals''---in essence, a
manual approximation of our approach.
Also in the context of the Equational Theories Project, Janota \cite{miko}
evaluated Vampire on the project's
problems and showed that
combining superposition with finite model finding can solve
almost all problems.
One reason for this success might be the work on term ordering diagrams by Hajdu et al.~\cite{hadju-et-al}.

We are aware of little work on automated proof minimization. Stachniak
\cite{Stachniak91} designed an algorithm for constructing resolution proofs
in propositional logics known as strongly finite logics.
Amjad \cite{Amjad07} and Cotton \cite{Cotton10} introduced techniques for minimizing
propositional resolution proofs.
Vysko\v{c}il et al.\ \cite{vyskocil-et-al} proposed to compress proofs by inventing new definitions using a heuristic based on \hbox{substitution} trees.
Gu et al.\ \cite{abs-2510-15700} developed Proof\-Optimizer,
which uses large language models to simplify Lean proofs.

Some SAT (satisfiability) and SMT (satisfiability modulo theories)
solvers can minimize the number of axioms needed for a proof, but the result
can be a longer proof.
SAT solvers commonly interleave search, which can be expressed
as resolution steps, with formula-rewriting techniques that go beyond resolution.
This interleaving, known as inprocessing~\cite{inprocessing}, is highly effective and
often yields both faster solving times and shorter proofs than either approach
in isolation.

The idea of automatically mixing and matching proofs is not new. Sutcliffe
et al.~\cite{Sutcliffe2011Combining} introduced a method for combining
automatically generated proofs to generate new ones.
Their proofs are represented as DAGs, enabling the identification and
replacement of
subproofs across different proofs. Proof combination is guided by heuristics
that measure structural similarity, and a greedy search strategy is used to
explore alternative combinations that yield proofs differing from the
originals.
In contrast to our approach, the main objective is to increase
proof diversity rather than minimize proof length.

\section{Conclusion}
\label{sec:conclusion}

Historically, more research has gone into finding proofs automatically than into
improving and presenting them.
We introduced an approach for minimizing equational proofs by mixing and matching
the output of separate runs of Vampire and Twee, and implemented it in
a new tool, Krympa. We used the tool to minimize the proof of problem
$650\Longrightarrow448$ from the Equational Theories Project from 62 to 20
steps, thereby providing a fully automatic solution to a challenge posed by
Tao. We also obtained remarkable reductions on other problems
originating from the project.
The shorter proofs are arguably easier to understand by humans and sometimes
more general. Our work
shows that proof automation and readability can go hand in hand.

\looseness=-1
Our approach could be extended in several ways.
First, it could be generalized to support full first- or higher-order logic.
Second, alternative lemma abstraction strategies could be explored.
Third, proofs with four or more segments could be synthesized.
Fourth, we might want to consider not only the number of steps but also
term size when measuring proofs, as suggested by Le Floch \cite{le-floch-tip}.
Fifth, we could try to translate Vampire proofs
to Twee's structured equality chain format.

Some possible extensions specifically concern the implementation.
First, we could explore whether nondefault Vampire and Twee
strategies can produce shorter proofs, following private suggestions by
Martin Suda and Nicholas Smallbone.
Second, since proof generation relies heavily on external provers,
performance could benefit from better scheduling of prover invocations, using
adaptive time limits.
Finally, our approach is not tied to Vampire and Twee specifically. Vampire is
representative of ATP systems that produce TSTP output, such as E and iProver,
whereas Twee is representative of completion-based provers such as Waldmeister.
Together, they cover most equational provers. Integrating further provers should
be straightforward.

\paragraph{\upshape\bfseries Acknowledgments.} We thank Bruno Le Floch,
Andrew Reynolds, Stephan Schulz, and Uwe Waldmann for fruitful discussions. We
thank Laura Kov\'acs, Luca Maio, Martin Suda, Mark Summerfield,
and the anonymous reviewers for helpful textual suggestions.

Blanchette's research was cofunded by the European Union (ERC, Nekoka,
101083038). Views and opinions expressed are however those of the authors only
and do not necessarily reflect those of the European Union or the European
Research Council. Neither the European Union nor the granting authority can be
held responsible for them.

Heule's research is supported by the NSF under grant DMS-2434625 and
funding from AFRL and DARPA under Agreement FA8750-24-9-1000.

\bibliographystyle{splncs04}

\begin{conf}
\bibliography{bibliography}
\end{conf}
\begin{rep}
\raggedright
\bibliography{bibliography}
\end{rep}
\end{document}